\documentclass{aa}
\usepackage{psfig}
\usepackage{amsmath}

        \newcommand{\as}{$^{\prime\prime}$}

\begin{document}
\title{Dust emission from the most distant quasars}

\author{F. Bertoldi\inst{1},
	C.L. Carilli\inst{2},
        P. Cox\inst{3}, 
        X. Fan\inst{4},
        M.A. Strauss\inst{5},
        A. Beelen\inst{3},
        A. Omont\inst{6},  
        R. Zylka\inst{7}
        }
\offprints{F. Bertoldi, bertoldi@mpifr-bonn.mpg.de}

\institute{Max-Planck-Institut f\"ur Radioastronomie, Auf dem H\"ugel 69,
           D-53121 Bonn, Germany
      \and National Radio Astronomy Observastory, P.O. Box,
           Socorro, NM~87801, USA
      \and Institut d'Astrophysique Spatiale, 
	   Universit\'e de Paris XI, F-91405 Orsay, France
      \and Steward Observatory, The University of Arizona, 
           Tucson AZ~85721, USA
      \and Princeton University Observatory, Princeton, NJ~08544, USA
      \and Institut d'Astrophysique de Paris, C.N.R.S., 
	   98b bd. Arago, F-75014 Paris, France
      \and IRAM, 300 Rue de la Piscine, 38406 St. Martin d'Heres, France
}

\date{Received date 26 March, 2003/ Accepted date ** 2003}

\titlerunning{Dust emission from $z>6$ quasars}
\authorrunning{F. Bertoldi et al.}

\abstract{We report observations of three SDSS $z>6$ QSOs at
250~GHz (1.2~mm) using the 117-channel Max-Planck Millimeter Bolometer
(MAMBO-2) array at the IRAM 30-meter telescope. J1148+5251
($z=6.41$) and J1048+4637 ($z=6.23$) were detected with 250 GHz flux densities
of $\rm 5.0 \pm 0.6 \, mJy$ and $\rm 3.0 \pm 0.4 \, mJy$,
respectively. J1630+4012 ($z=6.05$) was not detected with a
$3\sigma$ upper limit of 1.8~mJy. Upper flux density limits from VLA
observations at 43 GHz for J1148+5251 and J1048+4637 imply steeply
rising spectra, indicative of thermal infrared emission from warm
dust.  The far-infrared luminosities are estimated to be $\approx
10^{13} \, L_\odot$, and the dust masses $\approx 10^8 \, M_\odot$,
{   assuming Galactic dust properties}.
The presence of large amounts of dust in the highest redshift QSOs
indicates that {   dust formation must be rapid during the early
evolution of QSO host galaxies. Dust absorption may hinder the escape
of ionizing photons which reionize the intergalactic medium at this
early epoch.}
}

\maketitle

\sloppy

\section{Introduction}
\label{sec:Introduction}

The search for the most distant and early galaxies has become a
rapidly evolving field in extragalactic astronomy.  Optical imaging
and spectroscopic surveys (Palomar Sky Survey, PSS; Sloan Digital Sky
Survey, SDSS, York et al.\ 2000) have revealed a large number of QSOs
up to redshifts of 6.4 (Fan et al. 2001, 2003). {   About 150
high-redshift QSOs selected from these surveys were recently observed
at millimeter wavelengths, detecting thermal emission from one third
of them (Omont et al. 2001, 2003; Carilli et al. 2001a),
reaching out to a redshift of 5.5 (Bertoldi \& Cox 2002). Although
these optically bright QSOs give a somewhat biased view on the
relation between the formation of stars, massive black holes, and
galaxies in the early Universe, they complement especially at the
highest redshifts the blank field submillimeter imaging surveys, which
have uncovered a population of $z>2$ dust-obscured starburst galaxies,
likely to be spheroidal galaxies in their formation stages (Smail et
al.\ 1997; Hughes et al.\ 1998)}.


Recently, Fan et al.\ (2003) discovered three  QSOs at $z>6$ in the
SDSS, including J1148+5251 at $z=6.41$, the QSO with the highest known
redshift.  The spectra of these QSOs show the Gunn-Peterson quenching
of continuum emission blueward of $\rm Ly\alpha$, thus probing the end
of the reionization epoch of the Universe (White et al.\ 2003). These
sources provide an opportunity to study the growth of massive black
holes and their associated stellar populations at the end of the
``dark ages'', in the earliest epochs of luminous cosmic structure
formation.

In this Letter, we report the detection of 250~GHz (1.2~mm) continuum
emission from two of the SDSS $z>6$ QSOs and upper limits {to
their 43~GHz continuum emission} which confirm that the emission is
thermal dust radiation, thus enabling us to estimate
far-infrared luminosities and dust masses. Throughout this paper, we
adopt a $\Lambda$-cosmology with $H_0=71\rm~km~s^{-1}~Mpc^{-1}$,
$\Omega_\Lambda=0.73$ and $\Omega_m=0.27$ (Spergel et al. 2003).

\section{Observations}  
\label{sec:observations}

 
The millimeter continuum measurements were made in January and
February 2003 using the 117-channel MAMBO-2 array (Kreysa et al.\ 1999) 
at the IRAM 30 m telescope on Pico
Veleta (Spain). 
MAMBO-2 has a half power spectral bandwidth of 210 and 290 GHz
with an effective frequency
of 250 GHz. The beam size on the sky is 10.7 arcsec. The sources were
observed with a single channel  using the standard on-off
mode with the telescope secondary chopping in azimuth by
32$^{\prime\prime}$ at a rate of 2$\,$Hz. 
For  flux calibration a number of calibration sources were
observed, resulting in an estimated absolute flux uncertainty of $15\%$. 
The total 
on plus off target observing time was 51, 128, and 68 minutes, 
for J1148, J1048, and J1630, respectively.
The data were analyzed using the MOPSI software package.
Correlated noise was subtracted from each channel using the
weighted average signals from the surrounding channels.

J1148 was imaged with MAMBO-2 using the on-the-fly
mapping technique with chopping in azimuth by 42~arcsec. Skynoise was
subtracted and the double beam maps were combined through
shift-and-add.
Five maps of  one hour duration each were
combined for the final image, which is displayed as signal/noise
contours in Fig.~1.



Continuum observations at 43~GHz of J1148+5251 and J1048+4637 were
done using the VLA in the D configuration (max. baseline $=$ 1
km). The sources were observed for 4 hours and 0.7 hour,
respectively. Standard amplitude calibration was performed using
3C~286.  Fast switching phase calibration was employed using celestial
calibrators within 3$\rm ^o$ of the target sources.  The calibration
cycle time was 200 seconds, and the phase stability was excellent, with
typical changes in antenna-based phase solutions between calibration
scans $\rm < 10^o$ on the longest baselines.  Images were generated
using the deconvolution task IMAGR in AIPS, and CLEANed to residuals
of 1.5$\sigma$.  The Gaussian restoring CLEAN beam was $\sim
1^{\prime\prime}$ (FWHM).  The r.m.s. noise level in the final image
for J1048+4637 is 0.37~mJy/beam, and 0.11~mJy/beam for J1148+5251.

\begin{figure}[tbhp]
  \centerline{\psfig{figure={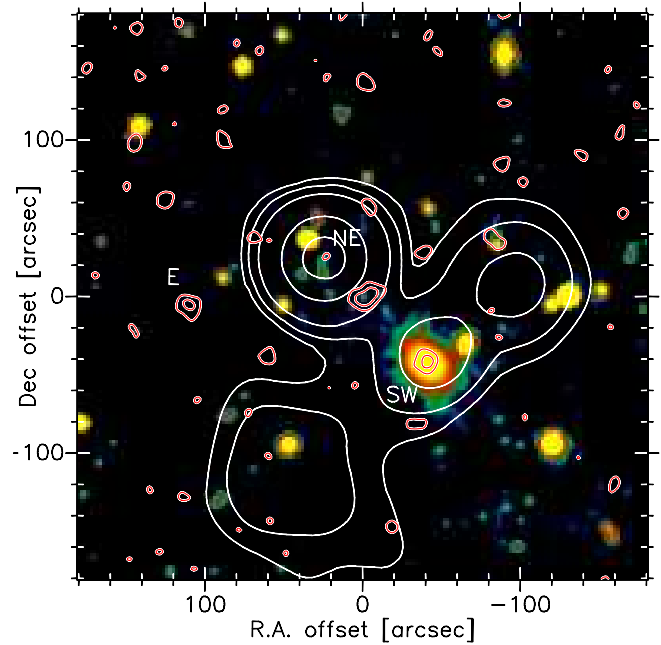},width=8.7cm,angle=-90}}
         \caption{ The 6 arcminute field surrounding the QSO SDSS
	   J1148+5251. Coordinates are offsets from the optical QSO,
	   which is visible as a faint blue dot.  {\em Color image:} SDSS
	   $z$ band image, smoothed to 5\as\ and logarithmically
	   scaled.  {\em Red-white contours:} MAMBO-2 250~GHz signal to
	   noise map smoothed to $13^{\prime\prime}$.  The contours
	   correspond to 2 and 4 $\sigma$.  The r.m.s. noise level,
	   $\sigma$, in the proper map is 0.9 mJy in the central
	   100\as\ and rises to $\sim 1.7$ mJy at a radius 200\as.
	   {\em White contours:} VLA NVSS 1.4 GHz image. The beam size
	   is $45^{\prime\prime}$, and contour values are 2, 4, 8, 32,
	   64 mJy/beam.  }
	 \label{fig:map} \end{figure}

\begin{table*}[tbhp]
\begin{center}
\caption{\small Properties of the observed QSOs and of field sources near J1148+5251\label{tab:qso}}
\begin{tabular}{lccccccccc}
\hline\\[-7pt]
Source & $z$ &  $M_{1450}$  & R.A.        &  Dec        & $S_{1.4}$    & $S_{43}$     &  $S_{250}$    & $L_{\rm FIR}$   & $M_{\rm dust}$ \\
        & & [mag]       &\multicolumn{2}{c}{(J2000)} & [mJy] & [mJy] & [mJy]  &  [$L_\odot$] & 
[$M_\odot$] \\[2pt]
\hline\\[-7pt]
J1630+4012 & 6.05 & $-26.1$ & 16 30 33.90 & $+$40 12 09.6 & $<$0.44     & --           & 0.8$\pm$0.6 & $< 5 \times 19^{12}$ & $< 2 \times 10^8$   \\
J1048+4637 & 6.23 & $-27.6$ & 10 48 45.05 & $+$46 37 18.3 & $<$0.43     & $<1.11$      & 3.0$\pm$0.4 & $7.5 \times 10^{12}$ & $4 \times 10^8$  \\
J1148+5251 & 6.41 & $-27.8$ & 11 48 16.64 & $+$52 51 50.3 & $<$0.33     & $<0.33$      & 5.0$\pm$0.6$^\dagger$  & $1.2 \times 10^{13}$ & $7 \times 10^8$  \\[3pt]
\hline\\[-7pt]
mm-SW      &      &         & 11 48 12.17 & $+$52 51 09   & $<$0.33     & $<$0.33      & 5.1$\pm$1.0   \\
FIRST-SW   & 0.05 &         & 11 48 12.16 & $+52$ 51 08   & 8.0$\pm$0.2$^{\dagger\dagger}$ & $<$0.33      &               \\[3pt]
mm-NE      &      &         & 11 48 19.30 & $+$52 52 14   & $<$0.33     & $<$0.33      & 2.2$\pm$1.0   \\ 
FIRST-NE   &      &         & 11 48 19.58 & $+$52 52 13   & 75$\pm$1    & 1.2$\pm$0.11 &               \\[3pt] 
mm-E       &      &         & 11 48 28.79 & $+$52 51 44   & $<$0.33     & $<$0.33      & 5.8$\pm$1.1   \\[2pt]
\hline
\end{tabular}
\end{center}
NOTE -- The optical properties are from 
        Fan et al. (2003) and the 1.4 GHz data from the VLA FIRST survey. 
        Upper limits are given at the $3 \sigma$ level. 
        $^\dagger$ Average of on-off and map
        measurements. 
        $^{\dagger\dagger}$ Source may be over-resolved in the FIRST survey.  \\  
        To estimate the  far-infrared luminosities, $L_{\rm FIR}$, we adopt a 
        dust temperature of 45~K and an emissivity index $\beta=1.5$, which 
        is typical for the spectrum of an infrared-luminous galaxy.
        Increasing $\beta$ to 2 would raise the luminosities by
        $\sim 20\%$, whereas varying the dust temperature by $\pm10$~K would
        change the luminosities by about a factor 2 down or up, respectively.
\end{table*}

\section{Results}
\label{sec:results}

The results from the MAMBO and VLA observations are summarized in
Table~\ref{tab:qso}. Two of the QSOs, J1148+5251 and J1048+4737,
are detected above $5 \sigma$ significance at 250~GHz (rest frame
165~$\rm \mu m$).  These QSOs are not detected at 43~GHz (rest frame
$970 \, \mu$m), with upper limit flux densities below 1~mJy. The steep
rest-frame submillimeter spectral index suggests that the emission is
thermal dust radiation and not synchrotron radiation.

\noindent { \bf J1148+5251.}~~~ {The highest redshift source yet
discovered ($z=6.41$), J1148+5251 is an extremely luminous QSO
powered by a massive black hole of $3 \times 10^9 \, M_\odot$
accreting close to its Eddington limit (Willott et al.\ 2003).} In a
$K^\prime$-band Keck image (Fan et al. 2003), the source ($K^\prime$ =
16.9) is unresolved and no other optical source with $K^\prime > 21$
is present within 10\as\ of the QSO, thereby ruling out strong
gravitational lensing on arcsecond scales.  J1148+5251 is detected at
250~GHz in the pointed measurements at $4.8\pm 0.8$ mJy, and in the
map at $5.3\pm 1.0$ mJy. It was not detected at 43~GHz with
the VLA to a 3$\sigma$ limit of 0.33~mJy, implying a lower limit to
the spectral index between 43 and 250~GHz of $+1.6$.

J1148+5251 was not detected at 1.4~GHz in the VLA FIRST survey (Becker
et al.\ 1995).  However, the survey shows two bright
radio sources within 1$^\prime$ of J1148+5251 {(Fig.~1 and Table 1).}  The one
located north-east of the QSO, FIRST-NE, is also detected (unresolved) at
43~GHz, implying a falling spectrum between 1.4 and 43~GHz of index
$-1.2$.  The other FIRST source (FIRST-SW) is located south-west of J1148+5251,
coincident with a large elliptical galaxy at redshift 0.05 (Fan et
al. 2003), and is not detected at 43~GHz.  
{   The lower resolution VLA 1.4 GHz NVSS 
and Westerbork 327 MHz Northern Sky Survey (WENSS)
show that FIRST-SW is located at the center of extended
radio emission reaching about 4 arcminutes SE-NW (Fig.\ 1).}
Note that for a 1$^\prime$
field one expects only 0.003 sources by chance with $S_{1.4} > 8$~mJy
(Fomalont et al. 2003), so the presence of several
such objects in the vicinity of the QSO is remarkable.

The MAMBO image of J1148+5251 reveals at least two other sources
(Fig.~\ref{fig:map}): {the $z=0.05$ elliptical galaxy (FIRST-SW) 
south-west of J1148+5251}, and a source toward the east (mm-E), 
with no optical counterpart in the SDSS
images. We also notice a 2~$\sigma$ peak which coincides with the
strong FIRST source north-east of J1148+5251, corresponding to
the fainter part of a double compact optical galaxy.  {   Several
potential millimeter sources at the 3 $\sigma$ level are found in
the 1.2~mm map, but considering the size of the map they are not very
significant. }

It is peculiar that the QSO is surrounded by two strong millimeter
sources.  The MAMBO deep field surveys (Bertoldi et al. 2000a,b; Carilli et al. 2001b) 
show an average surface density of sources with
flux density $\rm >4 \, mJy$ of 0.02 arcmin$^{-2}$. The probability to
find {   two such millimeter sources within a 2~arcmin radius from
the QSO} is only 6\%. The optical images show that the QSO falls into a region
with an overdensity of foreground galaxies surrounding the
$z=0.05$ elliptical galaxy. That the central cD galaxy of a cluster
can show noticeable submillimeter emission was pointed out by, e.g.,
Edge et al.\ (1999). The association of the QSO with a
millimeter-bright cD could thus be interpreted as due in part to a
lens amplification of the QSO by the cluster or the dark matter halo
associated with the elliptical.  However, the low redshift of the cD
or cluster would not produce a strong amplification.  The presence of
an intervening C{\sc iv} absorption system at $z=4.95$ (White et
al. 2003) {   hints at the possible existence of another possible lens,
but a high-redshift lens would not 
produce a large amplification either.}

It remains unclear whether the eastern millimeter source mm-E or the
strong radio source FIRST-NE are associated with the QSO or with the
foreground cluster.

{   If we assumed that the source mm-E is lens amplified
by a factor 2, then the chance probability to find such a source  
within $2^\prime$ of the QSO is of order unity.} Although the
statistical evidence is weak, the optical,  millimeter, and radio data hint
at a mild lens amplification toward J1148+5251, which would lower its
implied luminosity and dust mass. Alternatively,
they hint at a possible overdensity of objects near the $z=6.4$ QSO.

\noindent { \bf J1048+4637.}~~~ At $z=6.23$, this optically very
luminous BAL QSO is the third most distant quasar identified to date. 
There is no evidence for
arcsecond scale gravitational lensing of this source (Fan et
al. 2003).  J1048+4637 is detected at 250~GHz, but not  at
43~GHz (Table 1), 
and we find a lower limit to the spectral index between 43 and 250~GHz of
$+0.6$. 
This QSO was not detected at
1.4~GHz in the FIRST survey to a 3$\sigma$ upper limit of
0.43~mJy~beam$^{-1}$. There is, however, a possible detection of a faint,
0.46~mJy radio source 20\as\ west of the QSO.

\noindent { \bf J1630+4012.}~~~ This $z=6.05$
QSO is the optically faintest $z>5.7$ QSO found in the SDSS (Fan et al. 2003).
J1630+4012 is neither detected at 250~GHz nor at 1.4~GHz, and no radio
sources are found within 1$^\prime$ from the radio source to this
limit. 

\section{Discussion}
\label{sec:discussion}

All five QSOs known to date at $z \ge 6$ (Fan et al. 2001, 2003) were
observed at 250~GHz to r.m.s sensitivities $\sim 1$ mJy (Petric et
al.\ [2003] place upper limits for J1030+0524 and J1306+0356 at $z=6.28$ and 5.99,
respectively). The detection of two quasars reported in this Letter is
consistent with the 30\% detection fraction of QSOs in the redshift range $z=
2$ to 6 surveyed at 250~GHz to mJy sensitivities (Omont et al. 2001, 2003;
Carilli et al. 2001a). The fraction of optically luminous QSOs that
are also infrared luminous is therefore roughly constant with
redshift, out to the highest redshifts explored.  If the dominant dust
heating mechanism is radiation from young stars, the implied star
formation rates in J1048+4637 and J1148+5251 are $2000 \,
M_\odot \,\rm yr^{-1}$ and $3000\, M_\odot\rm \, yr^{-1}$,
respectively, comparable to what was derived for the $1.5 < z <
5.5$ QSOs detected at millimeter wavelengths.

In the optical spectra of the two $z>6$ QSOs detected at 250~GHz,
the Ly$\alpha$+[N{\sc v}] emission lines are relatively weak, in
contrast to the three non-detected QSOs, which show stronger and
sharper lines (Fan et al. 2001, 2003). This trend agrees with
the results of Omont et al. (1996) that luminous high redshift QSOs
with weak broad emission or broad absorption optical lines tend 
to have stronger millimeter emission.

From the infrared luminosities, we derive dust masses of $4 \times
10^8$ and $7 \times 10^8 \, M_\odot$ for J1048+4637 and J1148+5251,
respectively. Following Omont et al.\ (2001) we here
adopted a dust absorption coefficient at $\rm 230 \, \mu m$ of
$\kappa_{\rm 230} = 7.5 \, \rm cm^2 \, g^{-1}$, {    a value that applies
to a galactic dust composition and is unknown for
high redshift sources. The dust mass is also affected by the exact
temperature of the warm dust component and by the possible presence of
an additional cold dust component (see the discussion in Omont et al. 2001).
Despite these large uncertainties,} the
estimated dust masses are huge, implying a high abundance of heavy
elements at $z \approx 6$. This is consistent with the super-solar
metallicities found in the three QSOs discussed here (Fan et al. in
preparation),
and with the Fe/Mg abundance ratios near or above the
solar value measured in three other QSOs at $5.7 < z < 6.3$
(Freudling et al. 2003).

The presence of large amounts of dust at redshift 6.4 implies that
efficient dust formation took place between the corresponding cosmic
time and the epoch of early reionization ($z_r \approx 17 $, Kogut et
al.\ 2003), a time span of $\approx 0.7$~Gyr.  At a constant formation
rate this implies a {net} dust production rate of $\approx 1 \,
M_\odot \rm\, yr^{-1}$.  

A time span of 0.7\ Gyr is short by at least a
factor 2 to efficiently produce refractory grains in the quiescent
winds of low-mass ($M \le 8 \, M_\odot$) stars.
If the observed dust were the product of stellar processes, the
initial refractory dust enrichment might have occurred primarily through
dust condensation in supernova remnants, and perhaps in the winds of
high-mass ($M \ge 40 \, M_\odot$) stars, which are thought to have
dominated the early phases of star formation (e.g., Bromm \& Loeb
2003). The dust in the early Universe must then be composed of silicates and
perhaps oxides, since carbon dust is primarily formed from stars of
mass 2--5~$M_\odot$ (Dwek 1998) -- except if dust production in the
winds of high-mass stars was important (Todini \& Ferrara 2001).
{    For silicate and iron oxide dust the mass absorption coefficient
may take a higher value (Henning \& Muschke 1997) than that we
adopted to compute the dust mass, which may
therefore be overestimated. }

Elvis et al. (2002) proposed that dust may be produced in an outflow
from the broad line region of the AGN.  The dust content of QSOs
should then be roughly independent of redshift, which is consistent
with the similar FIR properties of QSOs from redshift 2 to beyond 6.
A difficulty with this model is that it requires pre-existing heavy
elements in the interstellar medium, and hence prior
star formation may be required regardless.  This mechanism can produce
up to $10^7 \, M_\odot$ of dust fairly readily, but the production of
the much larger amounts may be problematic.

{    Although the quasars we observed are extreme and rare objects hardly
representative of the dominant star forming galaxies in the early
Universe, they do show that early star
formation lead to a rapid metal and dust enrichment of the
interstellar medium.  Absorption by dust could have significantly
reduced the escape fraction of ionizing radiation from galaxies during
the epoch of reionization.}

\acknowledgements 

We thank the IRAM staff, and especially A.\ Weiss for their untiring
support. Compliments to E.\ Kreysa and his group for providing a 
great bolometer array. {    Many thanks to N.\ Mohan
and C.\ de Breuck for pointing out the NVSS and WENSS radio data, and
to the referee,
R.\ Ivison, for his constructive comments.}
MAS acknowledges support from NSF grant AST-0071091.
The VLA of the 
National Radio Astronomy Observatory is a facility of the
National Science Foundation, operated under cooperative agreement by 
Associated Univ. Inc. The Institute for Radioastronomy at Millimeter
Wavelengths (IRAM) is
funded by the German Max-Planck-Society, the French CNRS, and the
Spanish National Geographical Institute. 

\end{document}